\documentclass[11pt]{article}

\def\mytitle#1{\setcounter{equation}{0}
\setcounter{footnote}{0}
\begin{flushleft}\Large\textbf{#1}\end{flushleft}
\vspace{0.25cm}}
\def\myname#1{\leftline{{\large #1}}\vspace{-0.13cm}}
\def\myplace#1#2{\small\begin{flushleft}\textit{#1}\\
\texttt{#2}\end{flushleft}}

\def\myclassification#1{\small\noindent
Pacs no :
       #1\vspace{0.5cm}}
\usepackage{graphicx}
\begin{document}

\mytitle{Holographic Dark Energy Model: State Finder Parameters}

 \myname{Nairwita
Mazumder\footnote{nairwita15@gmail.com}} \vskip0.2cm
\myname{Ritabrata Biswas\footnote{biswas.ritabrata@gmail.com}}
\vskip0.2cm \myname{Subenoy
Chakraborty\footnote{schakraborty@math.jdvu.ac.in}} \vskip0.2cm
\myplace{Department of Mathematics, Jadavpur University,
Kolkata-700 032, India.} { }

\begin{abstract}
In this work, we have studied interacting holographic dark energy
model in the background of FRW model of the universe. The
interaction is chosen either in linear combination or in product
form of the matter densities for dark matter and dark energy. The
IR cut off for holographic dark energy is chosen as Ricci's length
scale or radius of the future event horizon. The analysis is done
using the state finder parameter and coincidence problem has been
graphically presented. Finally, universal thermodynamics has been
studied using state finder parameters.\\

\end{abstract}
\myclassification{98.80.Cq,95.35.+d,98.80.-k}
\section{Introduction}
The standard cosmology is put into a great challenge by exciting
observational evidences \cite{Riess1,Perlmutter1} in the last
decade. To incorporate the present accelerating phase of the
universe within the frame work of Einstein gravity, it is
inevitable to introduce a non-gravitating type of matter with
hugely negative pressure (of the order of its energy density)
called dark energy (DE). For the mysterious DE, there are only
very few constraints on its form of an equation of state
\cite{Perlmutter1,Percival1}. The general and leading choice of
the unknown DE is the cosmological constant ($\lambda CDM$ model)
which represents a vacuum energy density having constant equation
of state $\omega =-1$. However, its observed value is far below
than the estimation from quantum field theory (known as
Cosmological constant problem). Also there is no explanation why
the constant vacuum energy and matter energy densities are
precisely of the same order at present epoch (known as Coincidence
problem). Due to both theoretical \cite{Weinberg1,Copeland1} and
observational \cite{Peebles1} problems related to cosmological
constant other DE models (varies with time) are used in the
literature. Scalar field models \cite{Ratra1,Cladwell1,Copeland2}
(commonly known as quintessence) have attracted special
attention compare to other alternatives \cite{Copeland1}.\\

At present DE and cold dark matter (hereafter called as DM) are
dominant sources (about  70 percent of DE and 25 percent of DM) of
the content of the universe. So it is natural to have a lot of
interest in studying coupling in the dark sector components
\cite{Ma1,Pavon1,He1,Hooft1,Cohen1,Peebles1,Mangano1,Micheletti1,Sandro1}.
Also it is possible to have information about these components
through gravitational interaction. Further, recently, it has been
shown that an appropriate choice of the interaction between DE and
DM can alleviate the coincidence problem \cite{Pavon2,Olivares1,Chen1,Amendola1,Boehmer1}.\\

For the unknown and mysterious nature of DE, it it is possible to
have some insight demands that DE should be compatible with
holographic principle which states that "the number of relevant
degrees of freedom of a system dominated by gravity must vary
along with the area of the surface bounding the system"
\cite{Hooft1}. Such a DE model known as Holographic DE (HDE)
model. Further the energy density of any given region should be
bounded by that ascribed to a Schwarzschild  black hole (BH)
 that fills the same volume \cite{Cohen1}. If we denote the DE density
 by $\rho_D$, $L$- size of the region (infrared cut off) and
 $M_p={(8 \pi G)}^{\frac{-1}2}$ the reduced planck mass, then
 mathematically $\rho_D \leq M_p^2 L^{-2}$. as a standard practice
 we write DE density $\rho_D$ as $$\rho_D={3M_p^2c^2}{L^2}$$ where
 the dimensionless parameter $c^2$ takes care of the
 uncertainties of the theory and the factor of three has been
 introduced for mathematical convenience. In the present work we
 choose $L=R_{RC}$, the Ricci's length and $L=R_E$, the radius of
 the future event horizon as the IR-cut off in two different
 sections.\\

 The argument behind the choice of the Ricci's length as
 \cite{Gao1,Xu1,Suwa1,Lepe1}as IR cut off is that it corresponds
 to the size of the maximal perturbatio, leading to the formation
 of a black hole \cite{Brustein1}. On the other hand, radius of
 the future event horizon is commonly used as the IR cut off of
 HDE models which gives the correct equation of state and the
desired accelerating universe. However,
 recently it has been shown \cite{Duran1} that future event
 horizon suffers from a severe circularity problem.\\

 For the interacting DM and DE to resolve the coincidence problem
 (as mentioned above), the interaction term is chosen in the
 present work as (i) a linear combination of the energy densities
 of the two matter components i.e. of the form
 $3b^2H(\rho_m+\rho_D)$, (ii) a natural and physically viable
 interaction term of the form $\gamma \rho_m \rho_D$ with $\gamma$
 a dimensionally $\left(\frac{L^3}{mt}\right)$ constant. I both
 the choices the interaction term should be positive definite (i.e.
 $r>0$)so that there is transfer of energy from DE component to DM
 section. This choice is favorable to solve the coincidence
 problem. Also validity of the second law of thermodynamics and
 $Le Ch\hat{a}telierr's$ principle \cite{Pavon1,Lip1} demand
 positive interaction term. It should be noted that the second
 choice of interaction term gives the best fit to observations
 \cite{Pavon1,Lip1}for HDE models. Lastly, we have not included
 baryonic matter in the interaction due to the constraints imposed
 by local gravity measurements \cite{Ratra1,Lip1,Hagiwara1}.\\

 In 2003 Sahni et al \cite{Sahni1,Debnath1} proposed state-finder
 parameters $\{r,s\}$ which are defined as

 \begin{equation}\label{intro1}
 r=\frac{1}{aH^3}\frac{d^3a}{dt^3}~~ \mbox{and}~~s=\frac{r-1}{3(q-\frac{1}2)}
\end{equation}

where $a$ is the scale factor for the FRW model, $H$ and
$q(=-\frac{a \ddot{a}}{{\dot{a}}^2})$ are the Hubble parameter and
the deceleration parameter respectively. In fact the parameter
$'r'$ forms the next step in the hierarchy of geometrical
cosmological parameters after $H$ and $q$. These dimensionless
parameters characterize the properties of dark energy in a
model-independent manner. According to Sahni et al trajectories in
the $\{r,s\}$ plane plane corresponding to different cosmological
models demonstrate qualitative different behavior. According to
them, the state finder diagnostic together with SNAP observations
may discriminate between different DE models.\\

In the present work, we analyze interacting holographic dark
energy model in terms of state-finder parameters and possible
resolution of the coincidence problem has been presented
graphically. Finally, we have analyzed universal thermodynamics in
terms of $\{r,s\}-$parameters. The validity of generalized second
law of thermodynamics result in some summarize the whole work.

\section{Interacting Holographic DE model at Ricci scale}\label{chapter2}
In this section, we consider homogeneous, isotropic and spatially
flat FRW model of the universe bounded by Ricci's length
$R_L={(\dot{H}+2H^2)}^{\frac{-1}2}$. The Einstein field equations
are
\begin{equation}\label{1}
3H^2=\rho_m +\rho_D
\end{equation}
and
\begin{equation}\label{2}
2\dot{H}=-(\rho_m+\rho_D+p_D)
\end{equation}

where $\rho_m$ is the energy density of the dark energy in the
form of dust while $(\rho_D,p_D)$ are the energy density and
thermodynamic pressure of the dark energy in the form of a perfect
fluid having equation of state $p_D=\omega_D
\rho_D~(\omega_D,~a~variable)$. Due to holographic nature of the
DE, the energy density $\rho_D$ has the expression

\begin{equation}\label{3}
\rho_D=\frac{3c^2}{R_L^2} \end{equation}

where IR cut off is chosen at the Ricci's length. As a
consequence, the equation of state parameter becomes

\begin{equation}\label{4}
\omega_D=-\frac{2}{3c^2}+\frac{1}{3\omega_D}
\end{equation}
where $\Omega_D=\frac{\rho_D}{3H^2}$ is the density parameter for
the HDE.\\

To determine the evolution equation for the density parameter we
start with the energy conservation relations

\begin{equation}\label{5}
{\dot{\rho}}_m+3H\rho_m=Q
\end{equation}

and

\begin{equation}\label{6}
{\dot{\rho}}_D+3H\rho_D(1+\omega_D)=-Q
\end{equation}

First of all in absence of interaction (i.e. $Q=0$) $\Omega_D$
evolves according to

\begin{equation}\label{7}
{\dot{\Omega}}_D=H(1-\Omega_D)(1-\frac{2\Omega_D}{c^2})
\end{equation}
using $x=ln a$ we have $\frac{d}{dx}=\frac{1}H\frac{d}{dt}$ and
the above equation becomes

\begin{equation}\label{8}
\frac{d\Omega_D}{dx}=-(1-\Omega_D)(1-\frac{2\Omega_D}{c^2})
\end{equation}

which on integration gives
$$\Omega_D=\frac{e^{\frac{2x}{c^2}+2c_1}-c^2e^{x+c^2 c_1}}{e^{\frac{2x}{c^2}+2c_1}-2e^{x+c^2 c_1}}$$

where $c_1$ is the integration constant. Now we shall introduce interaction in the following ways:\\

{\bf Case I:}
\begin{equation}\label{9}Q=3b^2H(\rho_m+\rho_D)\end{equation}
Using both the conservations, the evolution of  $\Omega_D$ is
given by

\begin{equation}\label{10}
{\dot{\Omega}}_D=-3H\left[-\frac{2\Omega_D(1-\Omega_D)}{3c^2}+\frac{(1-\Omega_D)}3+b^2\right]
\end{equation}

or equivalently,

$$\frac{d \Omega_D}{dx}=-\left[(1-\Omega_D)(1-\frac{2\Omega_D}{c^2})+3b^2\right]$$
The solution gives

$$\Omega_D=\frac{1}4\left(2+c^2+K tan\left[\frac{1}2\left(-\frac{Kx}{c^2}+K c_2\right)\right]\right)$$

where $K=\sqrt{-4+4c^2+24b^2c^2-c^4}~and~c_2$ is the integration
constant.\\

{\bf Case II:} \begin{equation}\label{11}Q=\gamma \rho_m
\rho_D\end{equation}

This choice of the interaction term with the conservation
equations results the evolution equation for $\Omega_D$ as
\begin{equation}\label{12}
\frac{d
\Omega_D}{dx}=-(1-\Omega_D)\left[(1-\frac{2\Omega_D}{c^2})+3\gamma
H \Omega_D\right]
\end{equation}

The state finder parameter as introduced by (\ref{intro1}) can
have the expressions for the present model as

\begin{equation}\label{13}
q=-\frac{\ddot{a}}{aH^2}=\frac{1}2+\frac{3}2\frac{p}{\rho_T}
\end{equation}

\begin{equation}\label{14}
r=1+\frac{9}2(1+\frac{p}{\rho_T})\frac{\partial p}{\partial
\rho_T},~s=\left(1+\frac{\rho_T}p\right)\frac{\partial p}{\partial
\rho_T}
\end{equation}

where $\rho_T=\rho_m+\rho_D$ is the total energy density. Thus
eliminating $\frac{\partial p}{\partial \rho_T}$ we have the
relation between $r$ and $s$

\begin{equation}\label{15}
\frac{2(r-s)}{9s}=\frac{p}{\rho_T}=\Omega_D \omega_D
\end{equation}

i.e. $$2(r-1)=9s\left[\frac{1}3-\frac{2
\Omega_D}{3c^2}\right]=3s\left(1-\frac{2\Omega_D}{3c^2}\right)$$
\begin{figure}
~~~~~~~~~~~~~~~~~~~\includegraphics[height=2in,
width=2in]{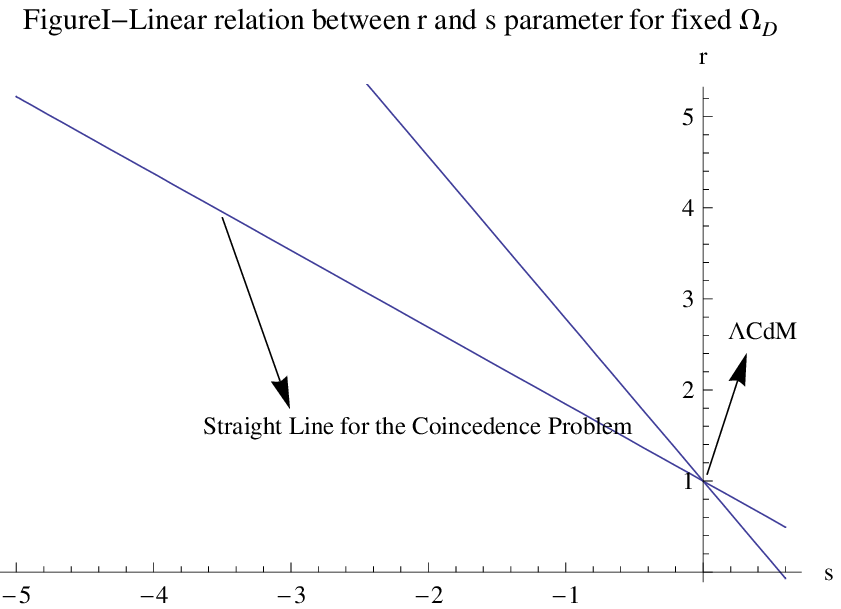}~~~~ \vspace{1mm} \vspace{6mm}
\end{figure}
So for fixed $\Omega_D$ we have straight line relation between the
two parameters $(r,s)$ i.e. we have a family of straight lines in
the $(r,s)-$plane with density parameter $\Omega_D$ as the
parameter. In fact trajectories in the $(r,s)-$plane correspond to
different cosmological models, for example the fixed point $(1,0)$
on the horizontal axis represents $\lambda CDM$ model. Further  as
$u=\frac{\rho_m}{\rho_D}=\frac{1}{\Omega_D}-1$ so from relation
(\ref{15}) the coincidence problem corresponds to the straight
line path $2(r-1)=3s(1-\frac{1}{c^2})$ in the $(r,s)$ plane.\\

\section{Interacting HDE model at future event horizon:}\label{chapter3}

The commonly used IR cut off for the holographic model is chosen
as the radius of the future event horizon $(R_E)$ given by

\begin{equation}\label{16}
R_E=a \int_t^{\infty}\frac{dt}a
\end{equation}

Normally, $R_E$ is chosen as IR cut off to have
correct accelerating universe.\\

Accordingly, the energy  density for the HDE can be written as

\begin{equation}\label{17}
\rho_D=\frac{3c^2}{R_E^2}
\end{equation}

For this choice of $\rho_D$ we shall now find the expressions for
the equation of state parameter $\omega_D$ and the evolution of
the density parameter $\Omega_D$.\\
For $Q=3b^2H(\rho_m+\rho_D)=3b^2H\rho_T$, we have

\begin{equation}\label{18}
\omega_D=-\frac{1}3
-\frac{2\sqrt{\Omega_D}}{3c}-\frac{b^2}{\Omega_D}
\end{equation}

and

\begin{equation}\label{19}
\frac{d
\Omega_D}{dx}=\Omega_D(1-\Omega_D)\left[1+\frac{\sqrt{2\Omega_D}}{c}-\frac{3b^2}{1-\Omega_D}\right]
\end{equation}

So from (\ref{15})

\begin{equation}\label{20}
2(r-1)=-3S
\left[\Omega_D+\frac{2{\Omega}_{D}^{\frac{3}2}}{c}+3b^2\right]
\end{equation}
\begin{figure}
~~~~~~~~~~~~~~~\includegraphics[height=2in,
width=2in]{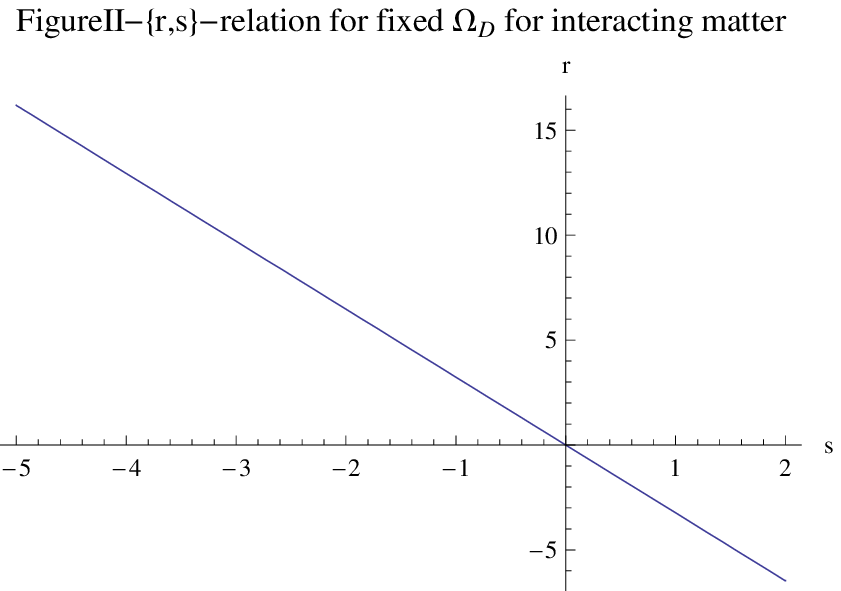}~~~~ \vspace{1mm} \vspace{6mm}
\end{figure}

So in this case the coincidence problem corresponds to the
straight line $2(r-1)=-3S(\frac{1}2+\frac{1}{\sqrt{2}c}+3b^2)$ in
$(r,s)-$plane. Similarly, for $Q=\gamma \rho_m \rho_D$ we obtain

\begin{equation}\label{21}
\omega_D=-\frac{1}3-\frac{2}3\frac{\sqrt{\Omega_D}}c-\gamma H
(1-\Omega_D)
\end{equation}
and
\begin{equation}\label{22}
\frac{d
\Omega_D}{dx}=\Omega_D(1-\Omega_D)\left[1+\frac{\sqrt{2\Omega_D}}{c}-3
\gamma H \Omega_D\right]
\end{equation}

Thus the $(r,s)-$ relation has the form

\begin{equation}
2(r-1)=-3S\left[\Omega_D+\frac{2{\Omega}_{D}^{\frac{3}2}}{c}+3
\gamma H (1-\Omega_D)\right]
\end{equation}

\section{Thermodynamics of the HDE model}

In this section, we study the thermodynamics of FRW universe
filled with interacting HDE. The universe is assumed to be bounded
by the horizon which we choose separately to be (i) apparent
horizon or (ii) event horizon or (iii) Ricci scale. In reference
\cite{Mazumder1} it has been shown that both the apparent and
event horizon do not change significantly over one hubble time
scale so it is reasonable to consider equilibrium thermodynamics
with temperature and entropy on the horizon similar to black
holes. If $S_I$ and $S_h$ denote the entropy of the matter
distribution inside the horizon and that of the horizon
respectively, then the time variation of the total entropy is
given by (for details see ref \cite{Mazumder1})
\begin{equation}\label{24}
\frac{d}{dt}{S_I+S_h}=\frac{4 \pi R_h^2
\rho_D}{T_h}\{u+(1+\omega_D)\}{\dot{R}}_h=\frac{12 \pi R_h^2
H^2}{T_h}\left[1+\frac{2(r-1)}{9s}\right]
\end{equation}
where $R_h$ is the radius of the horizon and $T_h$ is the
temperature of the horizon as well as of the inside matter for
equilibrium thermodynamics. To examine the validity of the
generalized second law of thermodynamics we first study the
evolution of the horizons.\\

For FRw model, the dynamical apparent horizon which is essentially
the marginally trapped surface with vanishing expansion, is
defined as a sphere of radius $R=R_A$ such that
$$h^{ab}\partial_a R \partial_b R=0$$ which on simplification gives

\begin{equation}\label{25}
R_A=\frac{1}{\sqrt{H^2+\frac{k}{a^2}}}
\end{equation}
The event horizon on the other hand is defined as Davis (1998) and
\cite{Mazumder1}

\begin{equation}\label{26}
\left.
\begin{array}{c}
~~~~~~~~~~~~~~-a sinh \tau,~k=-1 \\\\
R_E=-a \tau,~k=0\\\\
~~~~~~~~~~~~~~-a sin \tau,~k=+1
\end{array} \right\}
\end{equation}

where $\tau$ is the usual conformal time defined as,
$$\tau=-\int_t^{\infty}\frac{dt}{a(t)},$$

with $|\tau|<\infty$ for existence of event horizon. Also the
horizon radius corresponding to Ricci's length is given by

\begin{equation}\label{27}
R_L={(\dot{H}+2H^2)}^{\frac{-1}2}={(H\sqrt{1-q})}^{-1}
\end{equation}

where $q=-1-\frac{\dot{H}}{H^2}$ is the deceleration parameter. As
at present we are in a accelerating phase of the universe as the
event horizon exists and the horizons are related by the relations

\begin{equation}\label{28}
R_L<R_A=R_H<R_E,~ for~k-0
\end{equation}

where $R_H=\frac{1}H$ is the hubble horizon. The time variation of
the horizon radii are given by

\begin{equation}\label{29}
\left.
\begin{array}{c}
{\dot{R}}_A=\frac{H}2R_A^3 \rho_D[u+(1+\omega_D)] \\\\
{\dot{R}}_E=R_E\left[\frac{1}{R_A}-\frac{1}{R_E}\right]\\\\
{\dot{R}}_L=\frac{R_L^3H^3}2[(1+q)+(1-r)]
\end{array}
\right\}
\end{equation}
Usig the expressio for ${\dot{R}}_A$ from (\ref{29}) into
(\ref{24}) we obtain the standard result in the literature that
generalized second law of thermodynamics (GSLT) is valid both in
quintessence and in phantom era for the universe bounded by the
apparent horizon.\\

For the event horizon, as $R_E>R_A$ particularly in quintessence
era so that GSLT is satisfied unconditionally. On the other hand,
in phantom era we may have ${\dot{R}}_E>~or~<0$ (for details see
ref \cite{Mazumder1}, so the validity of GSLT is not
unconditional. In fact, as long as $\omega_D>-(1+u)~i.e.~s>0$ and
${\dot{R}}_E>0~i.e.~R_E>R_A$ then GSLT will be satisfied but for
$\omega_D<-(1+u)$ i.e. $s<\frac{2(1-r)}9$ we must have contraction
of the event horizon for the fulfillment of GSLT.\\

From Eq. (\ref{29}) we see that at the Ricci scale ${\dot{R}}_L$
increases until $r<(2+q)$ and GSLT will be satisfied provided
$-\frac{9s}2<r-1<1+q$. On the other hand, if the above inequality
is reserved i.e. $1+q<r-1<-\frac{9s}2$, though ${\dot{R}}_L$ is
negative and the expression within the curly bracket of Eq.
(\ref{24}) is negative but still GSLT will be satisfied. Further
as the state finder parameter $'r'$ is always greater than unity
and $'s'$ may have positive or negative values so the above
inequalities are modified as
$$r<2+q,~s>0~~or~~2+q<r<1-\frac{9s}2,~s<0$$ for the validity of
GSLT. These results are presented compactly in Table 1 and the
figure (III) shows the valid region in ${r,s}$-plane for
the validity of GSLT.\\\\
\begin{figure}
~~~~~~~~~~~\includegraphics[height=2in, width=2in]{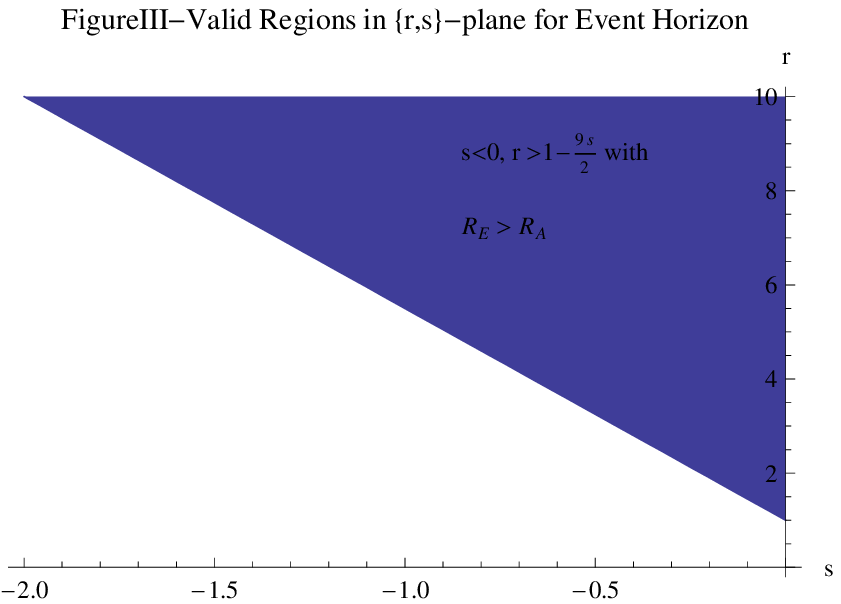}~~~~
\vspace{1mm} \vspace{6mm}
\end{figure}
{\bf Table I:Validity of GSLT in terms of $(r,s)$ parameters}\\\\
\begin{tabular}{|c|c|}

\hline\hline  Universe bounded by  & Condition for validity of GSLT
\\
\hline\hline Apparent Horizon  & Hold both in quintessence era and
in phantom era without any restriction.\\
\hline~~ Event Horizon~~  & Hold unconditionally in quintessence
era.
\\   & In phantom era : $R_{E}>R_{A}$ and $s<0$, $r>1-\frac{9s}2$
\\~~~~~ & or
\\~~~~~ &  $R_{E}<R_{A}$ and $s<0$, $r<1-\frac{9s}2$\\
\hline Ricci-scale length & Either $r<2+q$, $s>0$
\\~~~~~ &  or
\\~~~~~ &  $2+q<r<1-\frac{9s}2$, $s<0$\\

\hline
\end{tabular}\\\\\\

\section{Summary}

The paper deals with interacting holographic dark energy model in
the back ground of flat FRW model of the Universe. Two types of
interaction are chosen for investigation of which one is the
standard choice of the linear combination of the densities of the
two matter system. The other one which is physically reasonable
and supported by observational evidences is in the product form of
the energy densities. Here dark matter is chosen in the form of
dust while the HDE is chosen in the form of perfect fluid with
variable equation of state. The analysis is done for two choices
of the IR cut off of the HDE model namely Ricci's length scale and
the usual radius of the future event horizon. The state finder
parameters are introduced and the coincidence problem has been
presented graphically in the $(r,s)-$plane. Finally, universal
thermodynamics has been studied for this model with universe
bounded by the apparent horizon, or event horizon or by Ricci's
length scale. Validity of GSLT in all the three cases has been
showed in tabular form and the valid region in $(r,s)$ plane for
the validity of GSLT has been showed in figures $III$\\\\

\end{document}